\documentstyle[twoside,fleqn,espcrc2,fps]{article}

\newcommand{\AmS}{{\protect\the\textfont2
  A\kern-.1667em\lower.5ex\hbox{M}\kern-.125emS}}

\def\lsi{\raise0.3ex\hbox{$<$\kern-0.75em\raise-1.1ex\hbox{$\sim$}}}
\def\gsi{\raise0.3ex\hbox{$>$\kern-0.75em\raise-1.1ex\hbox{$\sim$}}}

\newcommand{\R}{{\kern+.25em\sf{R}\kern-.78em\sf{I} 
  \kern+.78em\kern-.25em}}

\newcommand {\beq}{\begin{equation}}
\newcommand {\eeq}{\end{equation}}
\newcommand {\beqa}{\begin{eqnarray}}
\newcommand {\eeqa}{\end{eqnarray}}
\newcommand {\n}{\nonumber \\}

\title{Maxwell Theory from Matrix Model}

\author{ Hiroyuki Takata\address{Institute of mathematical Sciences, 
C.I.T. Campus, Taramani, Chennai 600 113, India} 
\thanks{Email: takata@imsc.ernet.in. Talk presented at LATTICE 2000.}
}

\begin{document}

\begin{abstract}

We present a scenario for deriving Maxwell theory 
from IIB matrix model. 
Four dimensional spacetime  and theories on it relate different
dimensional ones by applying appropriate limits 
of the backgrounds of matrix model.
It is understood by looking at open strings bits 
as bi-local fields on the spacetime,
which are decoupled from the bulk in the limits.
The origin of electric-magnetic duality is also discussed in matrix model. 
\vspace*{-7mm}

\end{abstract}

\maketitle

\section{Introduction}
\setcounter{equation}{0}

It is necessary to make constructive definition of superstring theory
that derives well known physics. IIB matrix model\cite{IKKT} is a
candidate and it seems to  be suitable to the brane world
scenario\cite{RS}. 
It is defined in ten dimension and coordinates are
generalized to be noncommutative. We present the way to derive
commutative four dimensional theory from it.  


In IIB matrix model,
 we can find 3-brane classical solution of the action.
This 3-brane has a NCYM4 on it, 
where gauge fields are from the quantum fluctuation 
while spacetime is from the classical background.  
Since the 3-brane solution forms Heisenberg algebra,
4 dimensional space is constructed as  Hilbert space.
Although spacetime coordinate has uncertainty relation,
the coherent states make a intuitive 4 dimensional spacetime
as von Neumann lattice\cite{IKK}.
They showed open strings on the lattice 
and relates the noncommutativity to the string scale.  

There is the  correspondence 
between the  electromagnetic field
on the open strings
and the spacetime noncommutativity
in the matrix model\cite{KT}.
We are going to see the decouple limits in matrix model.
and get the intuitive explanation 
by using open strings bits on the von Neumann lattice. 
This is directly understood from relation 
between the noncommutativity $\theta^{01}$ from the matrix model 
and $\theta^{01}_{s}$ from string theory.
\section{Noncommutative U(1) theory with direction dependent noncommutativities}
\setcounter{equation}{0}

In this section
NCU(1) is derived from the matrix model.
In order to see the duality of the matrix model,
we first look into a theory around it, 
that is, NCU(1).

We start by following IIB matrix model action\cite{EK}\cite{IKKT}\cite{AIIKKT}:
\beq
S=- {1\over g^2}Tr({1\over4}[A_{\mu},A_{\nu}][A_{\mu},A_{\nu}]
+{1\over 2}\bar{\psi}\Gamma _{\mu}[A_{\mu},\psi ]) .
\label{IIBaction}
\eeq
Now $A_\mu$ and $\psi$ are $n\times n$ Hermitian matrices
and each component of $\psi$ is 10 dimensional Majorana-spinor.
We expand $A_\alpha=\hat{p}_\alpha+ \hat{\cal A}_\alpha$, 
(for $\alpha=0 \sim d$)around the following classical solution
\beq
\left[\hat{p}_\alpha,\hat{p}_\beta\right]
\eeq
\beq
=i \left( \begin{array}{ccccc}
0 & -1/\theta^{01} &0 &0& \\
1/\theta^{01} &0 &0 &0&\\
0 &0 &0 &-1/\theta^{23}& \\
0&0 & 1/\theta^{23}&0 &\\
&&&& \cdots
\end{array}
\right) ,
\label{sol}
\eeq
where $\theta^{01}$, $\theta^{23}, \cdots$ are c-numbers.
The noncommutative coordinates are introduced as:
\beq
\hat{x}^\alpha:=\theta^{\alpha \beta}\hat{p}_\beta\,\,,
\eeq
and satisfy the relation:
\beq
[\hat{x}^\alpha,\hat{x}^\beta]=-i \theta^{\alpha \beta}\,\,.
\label{xx}
\eeq
Since we are going to see different limits for different direction,

the duality and the decoupling limit,
not all $\theta$ are the same, 
namely, $\theta^{01} \not\equiv
\theta^{23}$ etc..

Followed Ref.\cite{AIIKKT}\cite{EMD},
$\hat{\phi}:=\{ \hat{\cal A}_\alpha, \hat{\varphi}_i:=A_i, \hat{\psi} \}$, 
($\alpha,\beta =0 \sim d, i,j = d+1 \sim 9 $)
are mapped to usual functions on phase space formed by noncommutative coordinates explicitly:
\beq
\hat{\phi} \rightarrow \phi(x)=\sum_k \tilde{\phi}_k e^{i k_\alpha x^\alpha}\,\,.
\eeq
The summation over $k_\alpha$ is performed as follows:
\beq
k_\alpha={l_\alpha \over n^{1 \over d} }\sqrt{2\pi \over |\theta^{\alpha,
\alpha+1}|}
\;,\,\,l_\alpha=\pm 1 ,\pm 2, \cdots, \pm {n^{2 \over d} \over 2}\,.
\label{momenta}
\eeq
Then we get the action of NCU(1) from eq.(\ref{IIBaction}):
\beqa
&&S_{\bf NCU(1)}={1\over g_{YM}^2}
\int d^{4}x
\Big({1\over 4}{\cal F}_{\alpha \beta}{\cal F}_{\alpha \beta}\n
&&+{1\over 2}[D_{\alpha},\varphi_i][D_{\alpha},\varphi_i]
-{1\over 4}[\varphi_i,\varphi_j][\varphi_i,\varphi_j]\n
&& -{i \over 2} \bar{\psi}{\Gamma}_{\alpha}[D_{\alpha},\psi ]
- {1 \over 2} \bar{\psi}\Gamma_i[\varphi_i,\psi ]
\Big)_{\star} .
\eeqa
Inside $(\mbox{\hspace{3mm}})_\star$, the products should be
understood as the star product:
\beqa
&&\phi_1(x)\star\phi_2(x)=\n
&&e^{{\theta^{\alpha \beta} \over 2 i}
{\partial^2 \over \partial \xi^{\alpha}\partial \eta^{\beta}} }
\phi_1(x + \xi) \phi_2(x+\eta)|_{\xi=\eta=0}
\label{star}
\eeqa 
The covariant derivative  and the field strength are defined as:
\beq
D_\alpha:=\partial_\alpha -i {\cal  A}_\alpha\,\,,\,\,\,\, 
{\cal F}_{\alpha\beta}:=i[D_{\alpha},D_{\beta}]_\star
\eeq

The Yang-Mills coupling is related to the noncommutativity as:
\beq
g^2_{YM}=g^2 (2\pi)^{d \over 2} \theta^{01}\theta^{23} \cdots \theta^{d-2\,d-1} \,\,.
\label{gYM}
\eeq
\section{Limits for four dimensional commutative Maxwell theory}
\setcounter{equation}{0}

In this section 
we are going to define a decoupling limit and
a commutative limit in the matrix model. 
We assume the spacetime dimension is almost equal to 4 
and coordinates are almost commutative. 
It dose not have to be exact 4 dimensional commutative spacetime.
The stand point of this paper is in 10 dimensional noncommutative one.  
Our strategy is getting the above brane world from the matrix model in 10 dimension,
by fine tuning the parameters $\theta^{\mu \nu}\,\,,(\mu, \nu =0 \sim 9)$.

To explain this by string terminology, 
we need identify strings in the matrix model.  
It is possible\cite{IKK}\cite{IIKK}.
We will summarize it in our case.
The von Neumann lattice is the best representing 
the intuitive spacetime.   
It is constructed by using coherent state 
of operators of the noncommutative coordinates
which forms Heisenberg algebra: eq(\ref{xx}). 
The lattice spacing is $\sqrt{2\pi \theta^{01}}$ for $0,1$ directions 
and  $\sqrt{2\pi \theta^{23}}$ for $2,3$ etc., 
which are written as $l^{\alpha}_{NC}$.
Because of the noncommutativity,  
states cannot be localized.
So, fields are naturally represented as bi-local ones,
which are functions of two points.
Small momentum modes
correspond ordinary (commutative) field.
Large momentum modes
correspond open strings. 
Define $d^\alpha:=\theta^{\alpha \beta} k_\beta$ 
and decompose $d$ as $d=d_0 + \delta d$, 
where $d_0$ is a vector 
which connects two points on the lattice and $|\delta d^\alpha| < l^{\alpha}_{NC}$.
The length of open string is $d_0^\alpha$
and the momentum which can be associated with the center of mass motion of open string 
is ${k_c}_\alpha:=(1/\theta)_{\alpha \beta} \delta^{\beta}d$. 
There is an inequality:
\beq
 |{k_c}_\alpha| < \sqrt{2\pi \over |\theta^{\alpha-1, \alpha}|}\,\,.
\eeq 

And the length and momenta are explicitly:
\beq
d_0^\alpha=m^\alpha \sqrt{2\pi \theta^\alpha}\,,
\hspace*{0.5cm}m^\alpha=0,\pm 1,\pm 2,\cdots,\pm {n^{1 \over d} \over 2}
\eeq

\beq
k_{c;\alpha}={m_\alpha \over n^{1 \over d}}\sqrt{2\pi \over \theta^\alpha}\,,
\hspace*{0.5cm}m_\alpha=0,\pm 1,\cdots,\pm (n^{1 \over d}-1)
\eeq

\subsection{Getting four dimensional spacetime}

We define the decoupling limit as:
\beq
\theta^{45,67,\cdots d-2\, d-1} \rightarrow \infty\,\,.
\label{decouple}
\eeq
In this limit, 

\beq
[\hat{p}_\alpha, \hat{p}_\beta ] \rightarrow 0\,\,\, \alpha, \beta
=4\sim d-1\,.
\eeq
This means the dimension of brane get down by two 
in $\alpha \beta$ directions, 
so it is natural definition.
The lengths and the momenta are
\beq
d_0^\alpha \rightarrow 0 \,\,\mbox{or}\,\,\, \infty 
\,\,\,,\,\,\,\,\, k_{c;\alpha} \rightarrow 0\,.
\eeq

The momenta of the open strings $k_c$ goes to zero.
If one considers the higher order correction to propergator of bi-local field,
then the oscillation of open string are seen 
by collecting open strings bits(see fig.).
In the limit, however, the momentum of the bits are goes to zero 
and the open string cannot make loop 
and is decoupled from closed strings in bulk.
$^{}$\footnote{
These features are parallel 
to string theory approach by the world sheet with NS-NS B filed\cite{ST}.
In order to see this clearly
we can use the relation 
between string theory and the matrix model. 
In the paper\cite{KT} we have gotten the relation 
between  IIB open string with a D-brane 
having both electric and magnetic field on the D-brane   
and the matrix model solution having both spacetime and spacespace noncommutativity.

Noncommutativities in matrix model $\theta^{01}$ 
and in string theory $\theta^{01}_s$ are
different from each other:
\beq
\theta^{01} \rightarrow \infty\,\,,\,\,\, \mbox{while}\,\, \theta^{01}_s: \mbox{finite}\,,
\eeq
and consistent with string theory approach.
}
\subsection{Commutative limit}

Let us think about the opposite limit:

\beq
\theta^{01,23} \rightarrow  0 
\,\,\,,\,\,\,\, n^{2 \over d} \theta^{01,23}: fix
\label{commutative}
\eeq

Since 
\beq
[x^\alpha ,x^\beta]_\star =-i\theta^{\alpha \beta}\,\,,
\eeq
coordinates commute to each other in this limit.
The momentum $k_\alpha$ is eigenvalue of $\hat{P}_\alpha 
:=[\hat{p}_\alpha, \mbox{\hspace{1mm}$\cdot$}\hspace{2mm}]$ 
(not of $\hat{p}_\alpha$).
Momenta commute to themselves without any limits 
because $[\hat{P}_\alpha, \hat{P}_\beta]=0$.
So this limit can be called commutative limits.

The lengths and momenta are

\beqa
d_0^{0\sim3}&=& 
0 \sim {1 \over 2}\sqrt{2 \pi n^{2 \over d} \theta^{01,23}} 
< \infty \n
&&\,\,\,\,\,\, (\#d=n^{1 \over d}+1)
\eeqa
\beqa
k_{c;0\sim3} &=& 
m_{0\sim3} \sqrt{2\pi \over n^{2 \over d} \theta^{01,23}} 
\leq \infty 
\,,\n
m_{0\sim3}&=&0, \pm 1, \cdots, \pm (n^{1 \over d}-1)
\eeqa

\vspace*{.5cm}

Now we can draw a scenario of getting an almost commutative near 4 dimensional spacetime.
We start from d dimensional solution of the matrix model.
There are $d/2$ noncommutativity parameters 
$\theta^{01}, \theta^{23}, \cdots, \theta^{d-2\,d-1}$.    
We  think of regions near following limits:
\beq
\theta^{01,23} \rightarrow 0 \,,\,\,
\theta^{45 \sim d-2\, d-1} \rightarrow \infty\,\,.
\eeq
Then we have a 4 dimensional commutative spacetime.
This is not dynamical determination, but just a fine tuning.

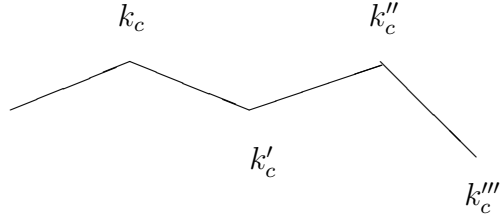
\begin{center}
\setlength{\unitlength}{3947sp}%
\begingroup\makeatletter\ifx\SetFigFont\undefined%
\gdef\SetFigFont#1#2#3#4#5{%
  \reset@font\fontsize{#1}{#2pt}%
  \fontfamily{#3}\fontseries{#4}\fontshape{#5}%
  \selectfont}%
\fi\endgroup%
\begin{picture}(2949,1260)(1489,-2311)
\thinlines
\put(1501,-1711){\line( 5, 2){750}}
\put(2251,-1411){\line( 5,-2){750}}
\put(3001,-1711){\line( 3, 1){832.500}}
\put(3826,-1411){\line( 1,-1){600}}
\put(2176,-1186){\makebox(0,0)[lb]{\smash{\SetFigFont{12}{14.4}{\rmdefault}{\mddefault}{\updefault}$k_c$}}}
\put(3001,-2086){\makebox(0,0)[lb]{\smash{\SetFigFont{12}{14.4}{\rmdefault}{\mddefault}{\updefault}$k_c^\prime $}}}
\put(3751,-1186){\makebox(0,0)[lb]{\smash{\SetFigFont{12}{14.4}{\rmdefault}{\mddefault}{\updefault}$k_c^{\prime\prime}$}}}
\put(4351,-2311){\makebox(0,0)[lb]{\smash{\SetFigFont{12}{14.4}{\rmdefault}{\mddefault}{\updefault}$k_c^{\prime\prime\prime}$}}}
\put(100, -2700){\makebox(0,0)[lb]{\smash{\SetFigFont{12}{14.4}{\rmdefault}{\mddefault}{\updefault}
$\mbox{ \hspace*{2cm}open string consisting of string bits.}$}}}
\end{picture}
\end{center}
\vspace*{.5cm}

\section{Electric-Magnetic Duality in Matrix Model (d=4 case)}
\setcounter{equation}{0}

In this section
we consider the electric-magnetic duality of NCU(1) 
on a D3-brane in the matrix model.

The relation between them are\cite{RU}:
\beqa
g_{YM D} &=& {1 \over g_{YM}}\,\,,\n
\theta^{\alpha \beta}_D
&=&{g_{YM}^2 \over 2}{\epsilon^{\alpha \beta}}_{\gamma \delta}\theta^{\rho \delta}\,\,,\n
{\cal F}_{\alpha \beta D}
&=&{1 \over 2 g_{YM}^2 }{\epsilon_{\alpha \beta}}^{\gamma \delta}{\cal F}_{\rho \delta} 
+ O(\theta)\,\,.\n
\label{dual0}
\eeqa
This is a spacetime-spacespace duality as well as electric-magnetic and strong-weak.

Next we will see this in our case.
By using eq.(\ref{gYM}), 
the dual Yang-Mills coupling and noncommutativity 
of eq.(\ref{dual0}) are written as:
\beqa
g^2_{YM D}&:=&{ 1 \over (2\pi g)^2 \theta^{01}\theta^{23}}\,\,,\n
\theta^{01}&:=&(2\pi g)^2 \theta^{01}(\theta^{23})^2\,\,,\n
\theta^{23}&:=&(2\pi g)^2 (\theta^{01})^2 \theta^{23}\,\,.
\label{dual}
\eeqa
Since there is a relation eq.(\ref{gYM}) in original theory, 
we would like the dual theory also to have the same one:
$g^2_{YM D}=(2\pi g_D)^2 \theta^{01}_D\theta^{23}_D$.
This requirement determine how the coupling $g$ changes to $g_D$:
\beq
g_D:={1 \over (2\pi)^4 g^3 (\theta^{01} \theta^{23})^2}
\label{dualg}
\eeq

We try to explain this by imaging a duality web.
The partition function of the matrix model is not changed 
under suitable rescaling of matrices.
That is, rescaling of coupling $g$ dose not change the model.
We have started with a $g$ 
and chosen an arbitrary back ground $\theta$.
On the other hand we can start by another $g_D$ and $\theta_D$,
and if those satisfy the condition:
\beq
(2\pi g)^2 \theta^{01}\theta^{23} \cdot (2\pi g_D)^2 \theta^{01}_D\theta^{23}_D=1\,\,,
\eeq
then two NCU(1)'s are the dual to each other.
This duality transformation is just rescaling: $g \rightarrow g_D$,
which is a symmetry.
It is natural to understand this 
if we remind type IIB superstring is self S-dual 
and its matrix model too.     

Finally let us see, in particular,  
more simple and familiar case.
We can find the dual pair of NCU(1)  from the matrix model with the same $g$,
and find the electric-magnetic duality for usual commutative Maxwell equations.
The U(1) coupling $g_{YM}$ cannot to be identity 
by rescaling of the gauge field ${\cal A}$ in noncommutative case. 
But, in the matrix model framework,  it is possible.
For given $g$, we choose the back ground solution with $\theta^{01}$ and $\theta^{23}$ which satisfy 
$(2\pi g)^2 \theta^{01}\theta^{23}=1$, namely, $g_{YM}=1$.
Then the dual noncommutativities also satisfy the same condition. 
Now the dual transformation is:
\beqa
(\theta^{01}_D,\theta^{23}_D) &=& (\theta^{23},\theta^{01})\,\,,\n
({\cal E}_D, {\cal B}_D ) &=& ({\cal B}, {\cal E} ) +O(\theta)\,\,.
\label{Maxdual}
\eeqa
Thus, Maxwell equations without the sources have a duality 
in  the following commutative limit: 
\beq
\theta, \theta_D \rightarrow 0\,\,,\,\,\, (2\pi g)^2 \theta^{01}\theta^{23}=1\,\,.
\eeq

\section{Conclusions}
\setcounter{equation}{0}

We have considered the decoupling-commutative limit 
and the electric-magnetic duality in the matrix model.
$d$dimensional spacetime can be constructed 
as D$(d-1)$-brane solution of the model.
It has non-selfdual solutions which we have treated in this paper. 
Then there are $d/2$ non commutativity parameters: 
$\theta^{01}, \theta^{23}, \cdots, \theta^{d-2\,d-1}$.
 
Electric-magnetic  duality  transformation  changes  those  parameters 
as eq.(\ref{dual}), as well as Yang-Mills coupling and electromagnetic fields. 
In addition, it corresponds to the rescaling of matrices 
in the original matrix model,
which has S-duality symmetry.
In particular solution related to a $g$, 
its duality is just the duality of U(1) Maxwell theory: eq.(\ref{Maxdual}).   

Decoupling limits have been defined as eq.(\ref{decouple}).
Open strings are decoupled from closed strings by looking into their momenta,
and the tension goes to zero.
This has been also seen from the relation $\theta^{01}$ and $\theta^{01}_s$ 
clearly
Commutative limits are defined as eq.(\ref{commutative}).
Noncommutativity parameters manage 
making our 4 dimensional commutative spacetime.
When $\theta^{45,67,\cdots,d-2\,d-1 } \rightarrow \infty$ 
corresponding directions decouple 
while $0 \sim 3$ direction commutes as  
$\theta^{01,23} \rightarrow 0$.
Staring from higher dimensional brane, 
we can get an  almost commutative and near 4 dimension spacetime,
by fine tuning of those parameters.

\end{document}